\documentclass[12pt]{iopart}

\usepackage{hyperref}
\usepackage{graphicx}
\usepackage{float}

\begin{document}

\title[First Principles Calculation of Field Emission]{First Principles Calculation of Field Emission from Carbon Nanotubes With Nitrogen and Boron Doping}

\author{Sherif Tawfik, Salah El Sheikh and Noha Salem}

\address{American University in Cairo}
\ead{shtawfic@aucegypt.edu}

\date{\today}

\begin{abstract}

We investigate the field emission properties of nitrogenated and boronated carbon nanotubes using time-dependent density functional theory, were the wave function propagation is performed using the Crank-Nicholson algorithm. We extract the current-voltage characteristics of the emitted electrons from nanotubes with different doping configurations. We found that boron doping either impedes, or slightly enhances, field emission. Nitrogen strongly influences the emission current, and the current is sensitive to the location of the nitrogen dopant in the nanotube. The emitted charge cloud from nitrogen doped carbon nanotubes is, however, more diffuse than that from pristine ones, our simulations show the emergence of a branching from the charge cloud, making nitrogenated carbon nanotubes less convenient for use in narrow beam applications.
\end{abstract}

\maketitle

\section{Introduction}

Field emission has been one of the active research areas owing to its theoretical, as well as commercial, significance \cite{TOhwaki}. Unique field emission properties were discovered in carbon nanotubes owing to their high aspect ratio and mechanical and chemical stability, making them the best candidate for flat-panel displays and atomic force microscopy probes \cite{YoshikazuNakayama,YSChoi,MMeyyappan}. Among the number of nanostructures investigated for field emission, carbon nanotubes proved, both theoretically and experimentally, to possess strongest field emission characteristics.

Nitrogen and boron are among the most common dopants encountered in carbon nanotubes \cite{SLatil}. Nitrogen-doped carbon nanotubes were studied experimentally by several groups \cite{SanjayK}\cite{SCRay}\cite{MayaDoytcheva}\cite{TJimbo}, reporting generally improved field emission characteristics. However, the strength of the emission current, as well as the onset field, were reported to be influenced by the substrate used in fabricating the nitrogenated carbon nanotube \cite{TJimbo}, processing method \cite{MayaDoytcheva} and post-processing \cite{SCRay}. While nitrogen doping of carbon nanotubes, as was indicated by, for example, Doytcheva \textit{et al.} \cite{MayaDoytcheva}, is close to that of metallic emitters, boron doping either impedes emission or has minimal effect on emission. Sharma \textit{et al.} \cite{Sharma2006} suggested that, based on their experimental investigation, boron doped carbon nanotubes are more convenient for applications in flat panel display devices where they are packed in large arrays or parallel emitters, while nitrogen-doped carbon nanotubes, owing to their high field emission quality, can be used in narrow beam applications.

Theoretical work by Qiao \textit{et al.} \cite{LQiao}, and experimental work by Chan \textit{et al.} \cite{LHChan} concluded that boron doping decreases the tunneling probability in comparison with pristine carbon nanotubes. Owens \cite{Owens2007}, based on theoretical computations, expected that both boron and nitrogen doping will not enhance field emission based on the fact that they increase the ionization energy of carbon nanotubes. However, although this is true in the case of boron doping, the ionization potential was not a correct indicator for field emission quality. Theoretical work on the field emission from boron doped carbon nanotubes, however, have reported conflicting results \cite{LQiao}, and there is a need to understand the details of the emission process and the variable factors to which emission current in boron doped carbon nanotubes would be sensitive. We believe that this work will shed more light into the microscopic nature of field emission from doped carbon nanotubes and the role of electron donation or acceptance in enhancing or impeding field emission.

\section{Computational Details}

We summarize our calculation procedure as follows (more details of the procedure are presented in Ref. \ref{SherifTawfik}). We performed Density Functional Theory computation using the Local Density Approximation for exchange and correlation potential by Pedrew and Zunger \cite{PedrewZunger}. Then, we solved the time-dependent Kohn-Sham equations by propagating the wave function using the Crank-Nicholson method. Then, following each propagation step, the amount of charge remaining in the emitter region at time $t$ for a particular wave function corresponds to the electric charge remaining, which is given by

\begin{equation}
\label{eq:Q}
Q_{n}(t)=\int^{z_{0}}_{0}\int\int\left|\psi\right|^{2}d\textbf{x},
\end{equation}

\noindent and in terms of the life time $\tau_{n}$ of state $n$,

\begin{equation}
\label{eq:Qfinal}
Q_{n}(t)=e^{-\frac{t}{\tau_{n}}},
\end{equation}

\noindent which implies linear behavior of $Q_{n}(t)$ in the short time interval. From this formula, the current generated by a wave function is given by

\begin{equation}
\label{eq:I}
I_{n}=e\frac{dQ_{n}(t)}{dt}\approx-\frac{1}{\tau_{n}}
\end{equation}

\noindent in the short time range. The total current is given by

\[
I=e\sum f_{n}\frac{dQ_{n}(t)}{dt}
\]

\noindent In order to compute the charge remaining in the lower portion of the simulation box, we integrate that lower portion each time-step for each wave function. At the beginning of the simulation, the charge is typically 1 (which works as a check for normalization consistency of the propagator used). Charge starts to decrease as time elapses, until it reaches a minimum point after which it starts increasing (due to reflection of the wave back from the upper surface of the simulation box). The integration is performed by simply summing the product of unit volume boxes of the mesh used in Octpus within the region concerned.

\section{Results and Discussion}

\par Here we consider three different nitrogen and boron doping configurations: the dopant replacing a carbon atom in the top ring; the dopant replacing a carbon atom in the second ring; the dopant replacing a carbon atom in the third ring. The different positions are shown in Fig. \ref{fig:cntn} below.

\begin{figure}	
	\centering
\includegraphics[width=50mm]{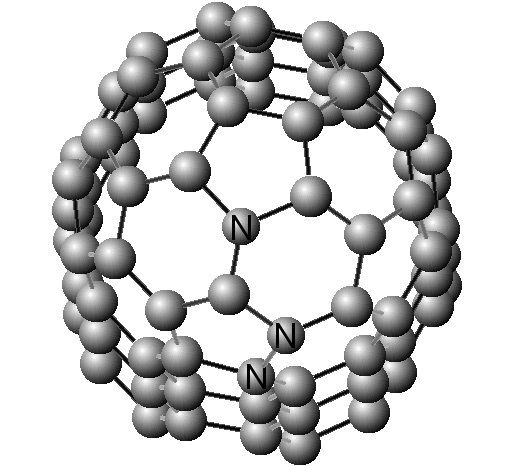}
\caption{Relative positions of nitrogen (or boron) dopant on the carbon nanotube tip}
\label{fig:cntn}
\end{figure}

\begin{center}
\begin{table}
\caption{Extracted current ($\rm\mu A$) from prestine vs. nitrogen-doped/boron-doped carbon nanotube against an applied electric field of $\rm 0.2 V/\AA$. All are H-passivated with no extra electrons. N1: Nitrogen dopant replacing a carbon atom at the top ring. N2: Nitrogen dopant replacing a carbon atom at the second ring below the tip. N3: Nitrogen dopant replacing a carbon atom at the third ring below the tip. B1: Boron dopant replacing a carbon atom at the top ring. B2: Boron dopant replacing a carbon atom at the second ring below the tip. B3: Boron dopant replacing a carbon atom at the third ring below the tip}
\label{tab:ncarbonnanotube}
	\begin{tabular}{| p{3cm} | c |}
		\hline
		Configuration & Current \\
		\hline
		Pure & 0.0161   \\
		N1 & 0.53394  \\
		N2 & 0.0796  \\
		N3 & 0.0164 \\
		B1 & 0.02198 \\
		B2 & 0.013469 \\
		B3 & 0.0268016 \\
		\hline
	\end{tabular}

	\end{table}
	\end{center}

\par
\begin{figure}[H]
	
	\centering
\includegraphics[width=100mm]{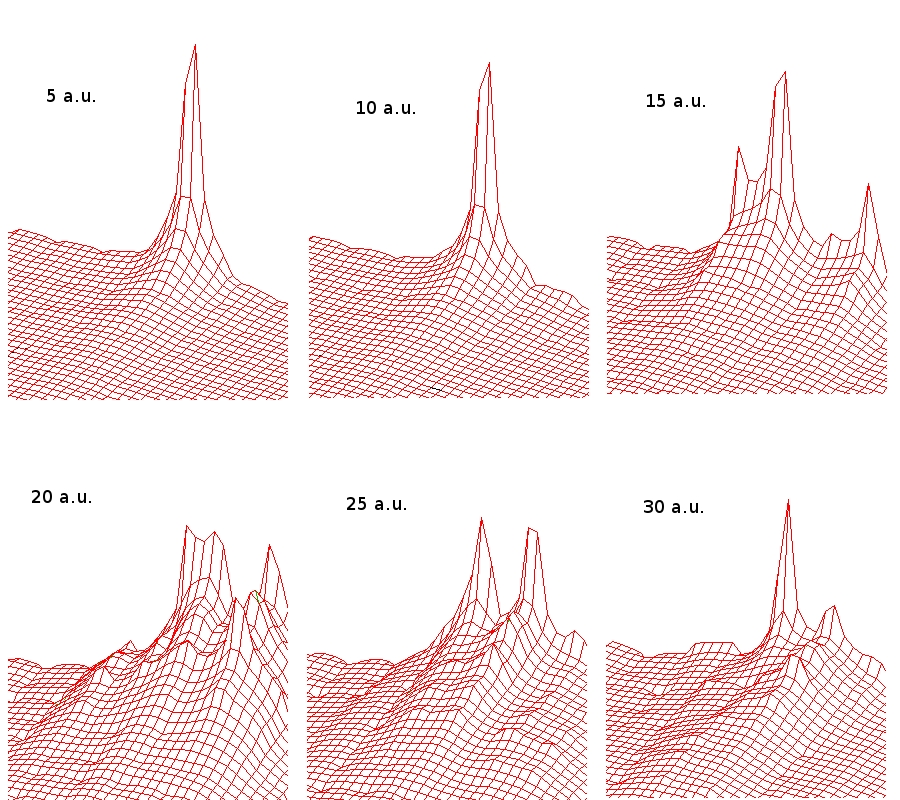}
\caption{Evolution of the wave function into the vacuum region in nitrogen-doped carbon nanotube: highest energy wave function. Units are arbitrary.}
\label{fig:N146upper}
\end{figure}

\begin{figure}[H]
	
	\centering
\includegraphics[width=100mm]{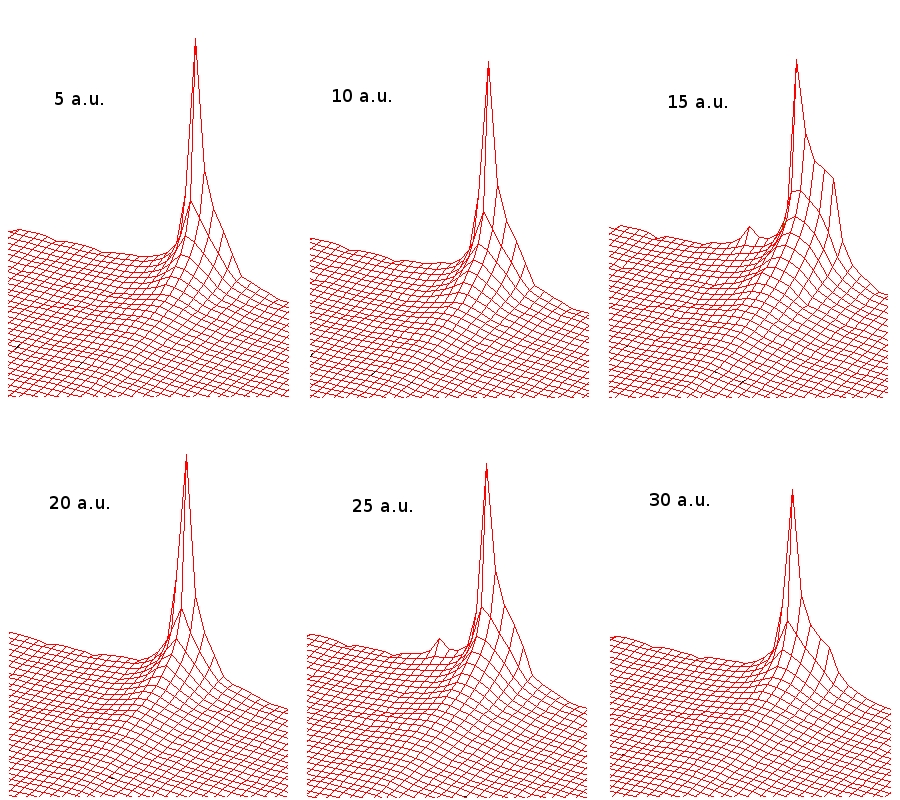}
\caption{Evolution of the wave function into the vacuum region in boron-doped carbon nanotube: highest energy wave function. Units are arbitrary.}
\label{fig:B145upper}
\end{figure}

 In order to understand the mechanism whereby a dopant enhances or impedes emission, we present in Fig. \ref{fig:N146upper,fig:B145upper} the charge evolution graphs for both the N1 and B2 configurations.

\begin{figure}[H]
	
	\centering
\includegraphics[width=100mm]{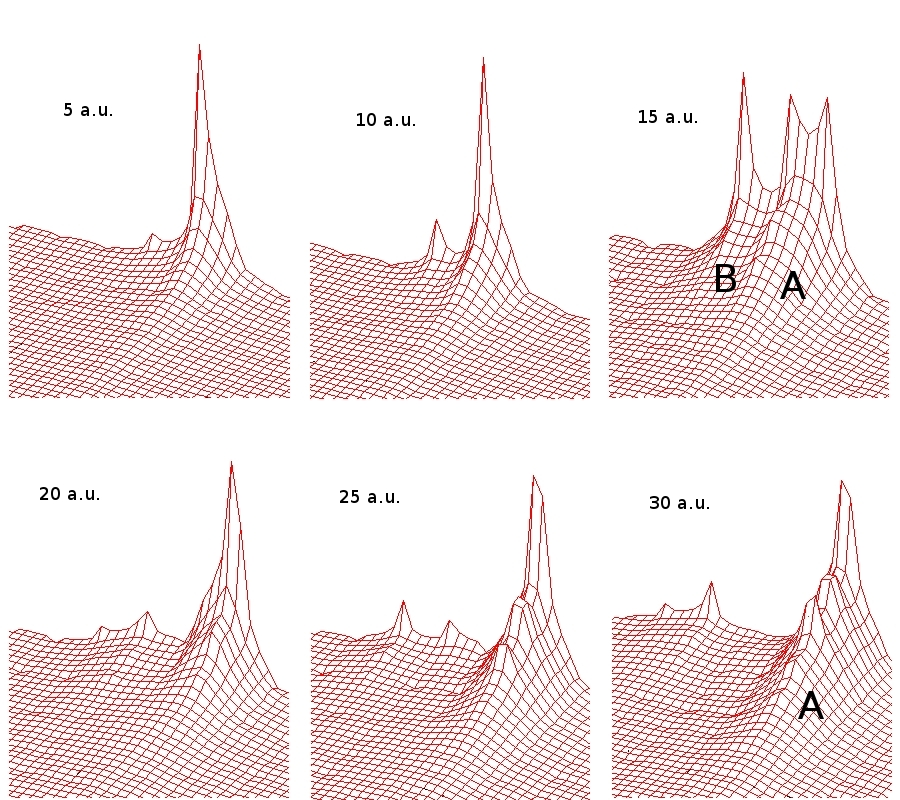}
\caption{Evolution of the wave function into the vacuum region in pristine carbon nanotube: highest energy wave function. Units are arbitrary.}
\label{fig:145upper}
\end{figure}

Fig. \ref{fig:N146upper} indicates the occurrence of a strong propagation of the wave function into the vacuum region. The influence of the nitrogen atom, however, did not concentrate the charge into a specific direction. Charge dispersed into a region that is almost $5\rm\mu \AA$ wide. Comparison with the charge evolution graph for pristine carbon in Fig. \ref{fig:145upper} shows the difference between the structure of the emitted charge cloud in pristine and nitrogen doped carbon nanotubes. In the former, the charge cloud tends to concentrate at specific regions, or channels, whereas in the latter case, the cloud tends to disperse in space.

It is an interesting observation that in nitrogen doping, the emission current is highly sensitive to the location of the dopant, which is not the case in boron doping. The location of a nitrogen atom at the uppermost ring (the tube's apex) results in a greatly enhanced emission current. As we place the nitrogen dopant further away from the tip, the emission enhancement drops drastically. This observation can be explained on the basis of the presence of the donated electron away from the tube body and closer to the tip region which already has increased electron localization. At the tube's apex, the donated electron enriches the heavy charge cloud in the tip region, and being away from the body of the tube, it experiences lower ionization potential. As the donated electron is placed further away from the apex, it experiences higher ionization potential imposed by the tube's body.

As for boron doping, results in Tab. \ref{tab:ncarbonnanotube} indicate that a boron dopant placed at the second top ring in the nanotube will impede emission, whereas placing it either at the apex or further down will enhance emission by increasing the emission current by 25\% compared to emission from a pristine carbon nanotube. Fig. \ref{fig:B145upper} below shows the impact of the boron dopant: the charge cloud seems to be almost fixed. This is due to the fact that boron is a charge acceptor; it attracts the electron cloud, contrary to the case of nitrogen doping.

\begin{figure}[H]
	
	\centering
\includegraphics[width=100mm]{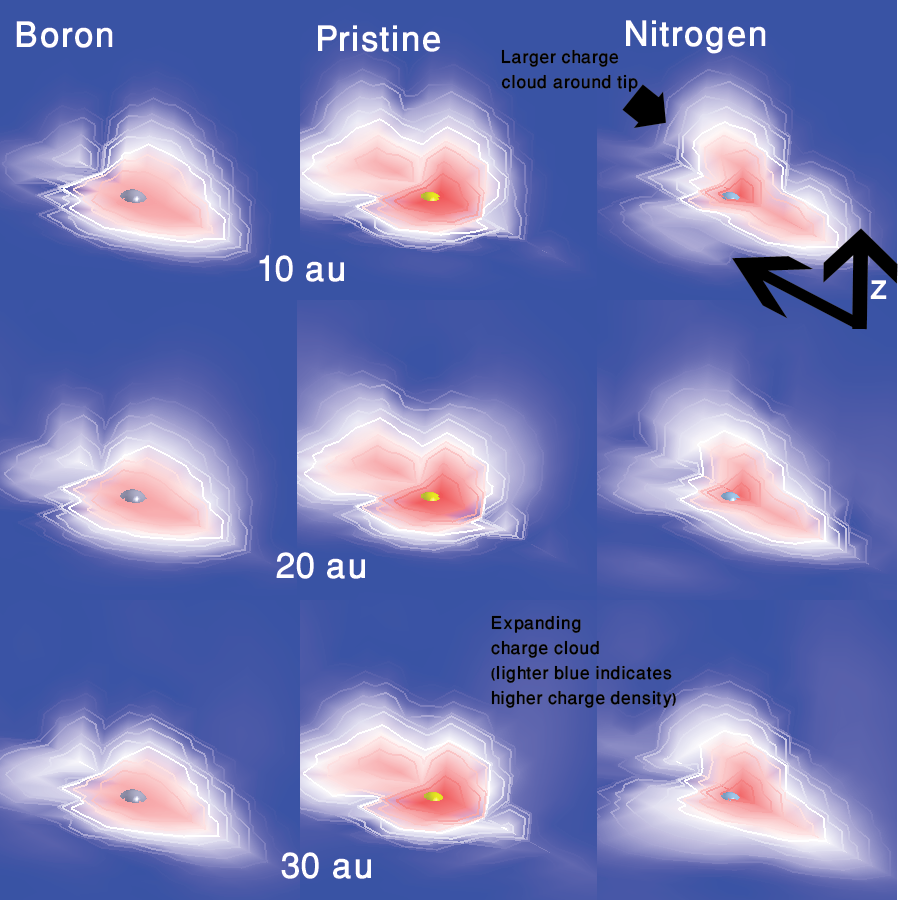}
\caption{Evolution of the wave function into the vacuum region in nitrogen-doped carbon nanotube: highest energy wave function. Units are arbitrary.}
\label{fig:comparison}
\end{figure}

Fig. \ref{fig:comparison} shows the difference between the charge cloud emerging in pristine, boron-doped and nitrogen-doped carbon nanotubes subject to $\rm 0.8 V/\AA$, where the dopant is placed at the nanotube's apex. Each column shows the evolution of field emission from the tip region of the nanotube after 10 au, 20 au and 30 au. The boron-doped carbon nanotube hardly emits charge. The extent of the charge cloud along the $x$ axis in boron is smaller than that in pristine, which is yet smaller than that in nitrogen-doped CNT. This is a consequence of the fact that boron is a charge acceptor, whereas nitrogen is a charge donor. Nitrogen doping clearly enhances the field emission by expanding the size of the charge cloud surrounding the tip region, and by liberating more charge as time elapses (which is depicted in the figure as a brightening of the blue color).

The extent to which the emerging charge cloud is diffuse in nitrogen-doped carbon nanotube can be inferred from Fig. \ref{fig:comparison-iso} below. Whereas the pristine carbon nanotube shows the emergence of a well-defined protrusion directed upwards, the charge cloud emerging from the nitrogen-doped nanotube diffuses to the surrounding tubes. Thus, although the signal from the nitrogen-doped nanotube is much stronger, it will be much harder to attempt to project it towards a particular target.

\begin{figure}[H]
	
	\centering
\includegraphics[width=100mm]{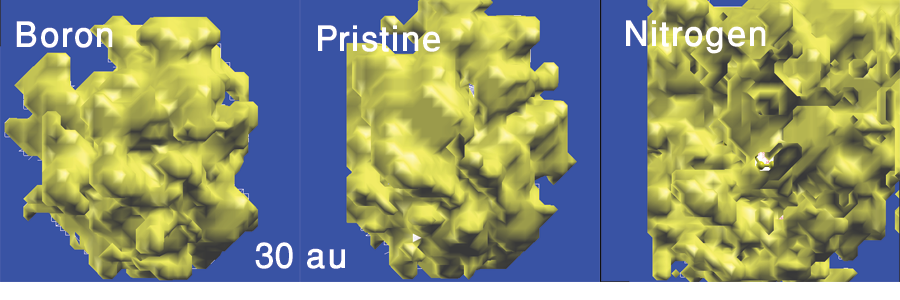}
\caption{Evolution of the wave function into the vacuum region in nitrogen-doped carbon nanotube: highest energy wave function. Units are arbitrary.}
\label{fig:comparison-iso}
\end{figure}

\section{Conclusion}

In summary, we have performed time-dependent density functional theory simulation of the field emission from nitrogen- and boron- doped carbon nanotubes. Boron tends to either weaken or slightly improve the emission current whereas nitrogen strengthens it. The strength of the emitted current in nitrogen-doped carbon nanotube is highly sensitive to the position of the dopant. And although the current emerging from the nitrogen-doped carbon nanotube is much stronger than that of pristine carbon nanotube, the charge cloud emerging from the latter is seen to project into a specific direction, unlike that of the former, which is diffused into space. We believe that the consideration raised in this paper about the extent of the sharpness of the emerging charge cloud will be of utility to applications of carbon nanotubes involving narrow electron beams, such as in Tunneling Electron Microscopy. Much work still needs to be done in identifying the dopants, as well as the conditions, that can produce narrower charge clouds.

\section{Acknowledgments}
We would like to thank the AUC Graduate Program for their support in this work.

\section{References}

\bibliographystyle{elsarticle/elsarticle-num}
\bibliography{mybib}

\end{document}